\documentclass[conference]{IEEEtran}

\usepackage{csquotes}
\usepackage{amsmath}
\usepackage{graphicx}
\usepackage{balance}
\usepackage[bookmarks=false]{hyperref}
\usepackage{caption}
\usepackage{subcaption}

\usepackage{cite}
\usepackage{url}
\begin{document}
\title{Multi-dimensional Conversation Analysis across Online Social Networks}

\author{\IEEEauthorblockN{William Lucia}
\IEEEauthorblockA{University of Insubria\\
Via Mazzini 5, Varese, Italy\\
w.lucia@studenti.uninsubria.it}
\and
\IEEEauthorblockN{Cuneyt Gurcan Akcora}
\IEEEauthorblockA{University of Insubria\\
Via Mazzini 5, Varese, Italy\\
cuneyt.akcora@uninsubria.it}
\and
\IEEEauthorblockN{Elena Ferrari}
\IEEEauthorblockA{University of Insubria\\
Via Mazzini 5, Varese, Italy\\
elena.ferrari@uninsubria.it
}}

\maketitle

\begin{abstract}

With the advance of the Internet, ordinary users have created multiple personal accounts on online social networks, and interactions among these social network users have recently been tagged with location information. In this work, we observe user interactions across two popular online social networks, Facebook and Twitter, and analyze which factors lead to retweet/like interactions for tweets/posts. In addition to the named entities, lexical  errors and expressed sentiments in these data items, we also consider the impact of shared user locations on user interactions. In particular, we show  that geolocations of users can greatly affect which social network post/tweet will be liked/ retweeted. We believe that the results of our analysis can help researchers to understand which social network content  will have better visibility.

\end{abstract}

\section{Introduction}

Recent years have seen an increasing number of successful online social networks catering to  different social needs on the Internet. From professional  social network LinkedIn to personal social  network Facebook, online social networks are diversified to meet needs of a growing online population. As a result, an Internet user has several accounts on online social networks, where his/her activities are governed by different motivations.

Two of the most successful online social networks, Facebook and Twitter, have grown in recent years to accommodate millions of social network users, and hold millions of personal profiles for the same sets of users. 
Dynamics of these online social networks are affected by different factors; the directed nature of Twitter enables fast and efficient information propagation, whereas undirected Facebook is still meshed by close familial or regional networks, and users interact in more conservative ways.

Despite these differences, both Facebook and Twitter provide means (i.e., like and retweet, respectively) to make another user's post visible to a larger audience. Analyzing how retweets and likes are employed provides clues in understanding how to spread ideas, disseminate news and propagate influence on a social network. 

 In this paper, we study the problem of how a post is liked/tweeted by social network users across Facebook and Twitter. Rather than external features such as the network structure or user information, we focus on tweets/posts themselves to understand what aspects of tweets/posts help them to get retweeted and liked. Moreover, we analyze how location information influences retweet, comment and like interactions.

To this end, we extract the following features from text based social network posts: named entities, lexical errors and sentiments. In named entity recognition, we locate atomic elements of seven categories: people, organizations, locations, date, time, percentage and money.
 By observing found entities, we show how the functionality (i.e., usage purpose) of online social networks can be mined. 

In lexical analysis, we locate lexical errors of texts, and classify these errors into ten most frequently committed errors. An analysis of these lexical errors show that social network users make similar mistakes on both sites. 

Sentiment analysis aims at studying user interactions for different sentiments that are found in social network posts. We investigate the impact of positive/negative/neutral sentiments on conversation patterns.

Furthermore, using location information from Twitter bios and Facebook profiles, we show that users' interaction patterns are heavily dependent on their current locations. Although online social networks have connected millions of users from all over the world, interaction patterns are still location limited, and governed by densely connected networks.

With the increasing popularity of online social networks, many studies  have been carried on  to better understand how humans communicate in a global setting (see e.g., \cite{ellison2007social,ye2010measuring}). For instance, a comprehensive analysis has found that online social networks, such as Twitter, can also act as an information source \cite{kwak2010twitter} to a high degree. This functionality of Twitter is provided by tweets and conversations that occur among users \cite{java2007we,nagarajan2010qualitative,suh2010want}. In conversational studies, the use of hashtags (i.e., Twitter topics) and addressivity (i.e., @ sign) have been found to increase interactions among Twitter users \cite{honey2009beyond}. However, in such work the studied dimensions are limited to  few features, such as hashtags, whereas we consider additional conversational dimensions such as sentiment analysis. Moreover, their analysis is limited to Twitter, whereas we focus also on Facebook posts.

From a theoretical point of view, sentiment analysis is a widely studied problem in research work \cite{kouloumpis2011twitter,agarwal2011sentiment}, but its impact on conversational interactions has not been well studied. 

Another dimension of our work is related to geolocation studies. Recent works by  Takhteyevet al. \cite{takhteyev2012geography} and Leetaru et al. \cite{FM4366} have analyzed the geography of Twitter and found that users communicate more often with those closest to them. However, these works are limited to Twitter only. In geolocation research, locations of friends have been studied from a privacy point of view on Facebook \cite{labitzke2013online}. In this work, we come to the similar conclusion that friends/followers share common locations to a high degree, but our focus is different in that we analyze the impact of locations on conversational interactions, by looking at 
like/retweet locations.

The paper is organized as follows. Section \ref{sec:datacollection} explains our data collection process on Facebook and Twitter.  Section \ref{sec:methodologytools} explains our methodology, and discusses the software tools we used. In Section \ref{sec:miningdimensions}, we analyze the four dimensions we considered, whereas Section \ref{sec:dimensioninterplay} shows how considered dimensions affect each other. Finally, Section \ref{sec:influencegeolocationconversations} combines our insights with geolocation data, and presents our results in a visual way.

\section{Data collection}\label{sec:datacollection}

Data sets of this paper have been created by querying Twitter.com and Facebook.com APIs.

A Facebook application \footnote{\url{developers.facebook.com}} was used to query Facebook for user data in an offline manner and store posts from the 2008-2013 period. Through the application, 75 users have given us permission to track data items (i.e., status posts, photos, videos, etc.)  along with comments and likes these items receive from other Facebook users. This data crawling allowed us to track conversations of 670K Facebook users. Each considered data item has been posted on Facebook by users, their friends or friends of friends. Similarly, likes/comments on these data items belong to the users, their friends or friends of friends. Facebook status posts can contain pictures and videos. As we cannot analyze contents of these additional data items, we limit our analysis to Facebook comments that consist of textual data only. In what follows, we will refer to Facebook comments as Facebook texts. 

Our Twitter dataset comes from a crawl between December 2012 and April 2013, and the considered tweets have been posted between 2008 and 2013. By using a Twitter application \footnote{\url{dev.twitter.com}}, we have stored bio information and last 10 tweets of 11M Twitter users. 

In what follows, we will use the terms Twitter texts and tweets interchangeably. For geolocation analysis we used bio information on user profiles. Bio information for each user contains a current location field, where unstructured texts can be entered for city/country values, e.g., Boston, MA. We have used Google Geocode API \footnote{\url{developers.google.com/maps/documentation/geocoding/}} to convert these location texts into longitude-latitude values.

\section{Methodology and tools}\label{sec:methodologytools}

In order to understand what factors improve the chances of a text getting liked/retweeted by other users, we have considered the following dimensions:  named entity recognition, lexical and sentiment analyses. In what follows,  we will  discuss our methodology for mining each of these dimensions.

\bigskip

In given sentences, named entity recognition (NER) \cite{nadeau2007survey} locates atomic elements of predefined categories such as names of people, cities, locations, time mentions, and money values. For  entity recognition, we used Apache OpenNLP \footnote{\url{http://opennlp.apache.org/}} that allows training its classifier with seven different categories. We have used the categories \textquote{people, organizations, locations, date, time, percentage and money}. This tool has a precision of 0.8 and a recall of 0.74.

\medskip

Lexical errors are grammar and spelling mistakes/typos found in sentences. These are noun-verb agreement errors, missing words, extra words, wrong words, confusion of similar words, wrong word order, comma errors, and whitespace errors.
For Facebook and Twitter texts, we used  LanguageTool\footnote{\url{http://www.languagetool.org}} \\to find lexical errors. These errors are predefined text patterns defined in an XML file. The tool reaches a precision of 93\% on the English texts.

\medskip

Sentiment analysis aims at extracting the general sentiment from a given sentence, identifying whether it expresses a positive, negative or neutral emotion.
 For this purpose, we used the \emph{Sentiment140 API}, a Sentiment Analysis tool developed by Go et al. \cite{go2009twitter}, which is a recent and widely used tool. The tool is based on a machine learning algorithm for classifying the sentiment of Twitter messages using distant supervision. The training data consists of Twitter messages with emoticons (i.e., pictorial representation of a facial expression), which are used as noisy labels. This type of
training data is abundantly available and can be obtained through automated means.  The underlying idea is to use emoticons to learn which words co-appear with emoticons, and use this information in machine learning algorithms, such as naive bayes and support vector machines. They show that the tool has an accuracy above 80\% when trained with emoticon data \cite{go2009twitter}. 

\section{Mining Dimensions}\label{sec:miningdimensions}

In mining the considered dimensions, we have focused on   English language texts. To this end, we have used the Ldig library in Python \footnote{\url{ https://github.com/shuyo/ldig}}, which has a precision of 99.1\% in detecting the English language.

For Facebook and Twitter datasets, counts of English language texts and dimensional statistics are given in Table \ref{table:dimstats}. In columns named entity, error and sentiment, we give the percentages of English texts which contain at least one named entity, lexical error and sentiment tag, respectively. In the following sections, these values will be explained in detail. 

\begin{table}
\centering
	\caption{Dimensional statistics and counts of English texts. Entitity, error and sentiment values are given in percentages.}
	\label{table:dimstats}
	\begin{small}
  \begin{tabular}{c|c|c|c|c|c|c|c|c|}
				\textit{} &Count & Entity & Error & Sentiment \\ 
				Twitter & 10.6M  & 29.8\%& 81\%& 32\%\\
				Facebook  & 1M  & 13\%& 69\%& 22\%\\ 
	\end{tabular}
		\end{small} 
\end{table}

\begin{figure*}[ht]
        \centering
        \begin{subfigure}[b]{0.4\textwidth}
                \centering
                \includegraphics[scale=0.7]{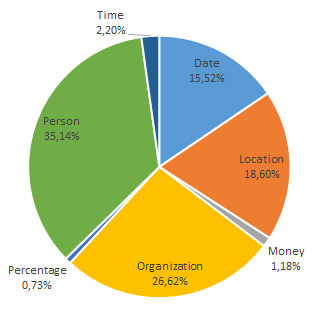}
				\caption{Facebook entities.}
                \label{fig:fbentities}
        \end{subfigure}%
       \begin{subfigure}[b]{0.4\textwidth}
                \centering
               \includegraphics[scale=0.7]{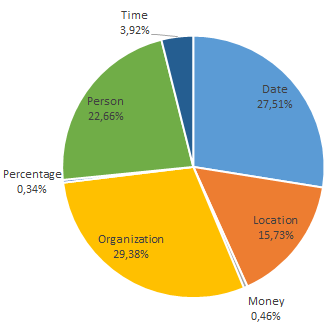}
				\caption{Twitter entities.}
                \label{fig:twentities}
        \end{subfigure}
         \caption{Entities and their percentage in Facebook and Twitter posts.}\label{fig:mixentities}
\end{figure*}

\subsection{Entity Recognition}

In Table \ref{table:dimstats}, percentage of English texts which contain at least one named entity are 29.8\% and 13\% for Twitter and Facebook texts, respectively. Figure \ref{fig:mixentities} further details the composition of entities in these texts. Along seven different categories, we see that on Facebook, Persons are mentioned in texts 35.14\% of the time, whereas this value is 22.66\% for tweets. Date and time entities are mentioned twice as much on Twitter, but organizations have similar percentage values. We attribute the difference in date/time values to Twitter users' high mobility (i.e., mobile phone usage) compared to other social network users \cite{lenhart2009twitter}. Because of this mobility, tweets are more related to events happening in real time. For example, 1.5\% of tweets contain the word today, whereas this value is only 0.02\% for Facebook texts. In time entities, we found a similar pattern with tonight appearing in 0.06\% of tweets and 0.0012\% of Facebook texts, respectively.

\begin{figure}[!ht]
\centering
\includegraphics[scale=0.5]{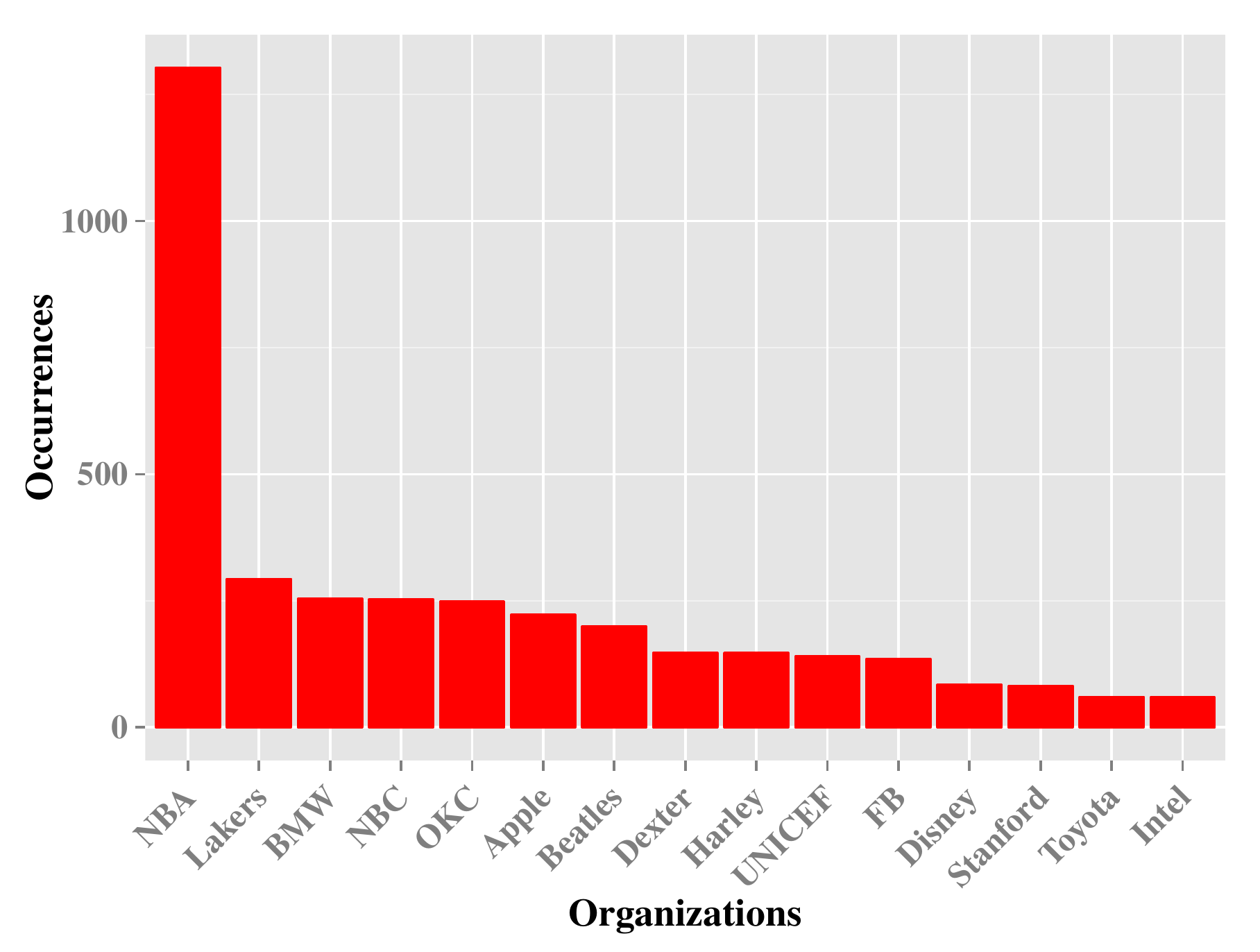}
\caption{Most frequently mentioned organizations on Facebook.}
\label{fig:fborg}
\end{figure}

\begin{figure}[ht]
\centering
\includegraphics[scale=0.5]{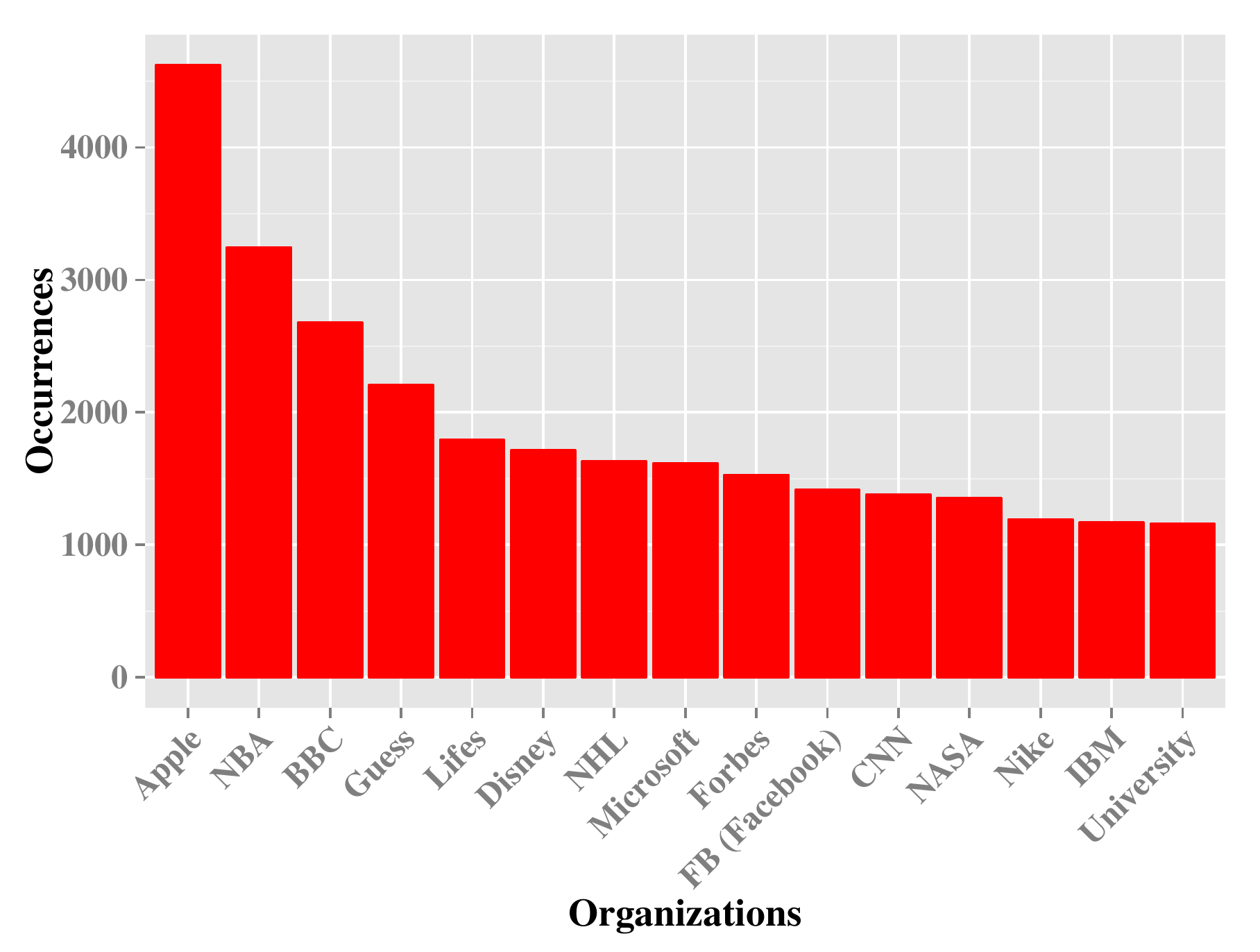}
\caption{Most frequently mentioned organizations on Twitter.}
\label{fig:tworg}
\end{figure}

As seen on Figures  \ref{fig:fborg} and \ref{fig:tworg}, some organizations are frequently mentioned on both Twitter and Facebook, but users on Twitter are more likely to mention a broader variety of organizations, ranging from commercial brands to news agencies. In contrast, the two most frequently mentioned organizations on Facebook are related to  basketball, and other organizations are not very frequently mentioned.

\begin{figure*}[ht]
\centering
         \begin{subfigure}[b]{0.4\textwidth}
				\centering               
               \includegraphics[scale=0.45]{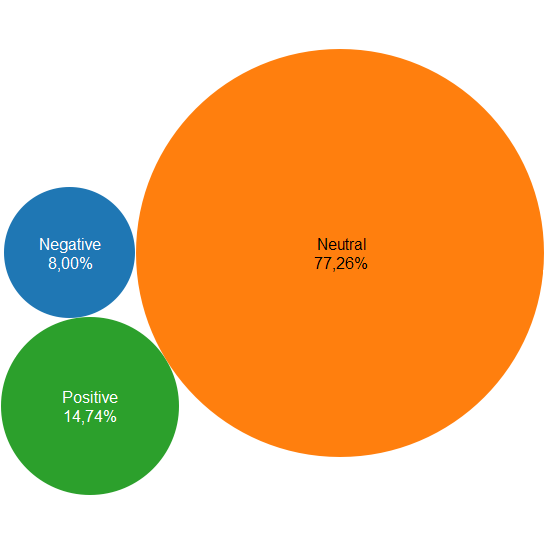}
				\caption{Facebook sentiment}
                \label{fig:fbsentiment}
        \end{subfigure}%
       \begin{subfigure}[b]{0.4\textwidth}
       			\centering
               \includegraphics[scale=0.45]{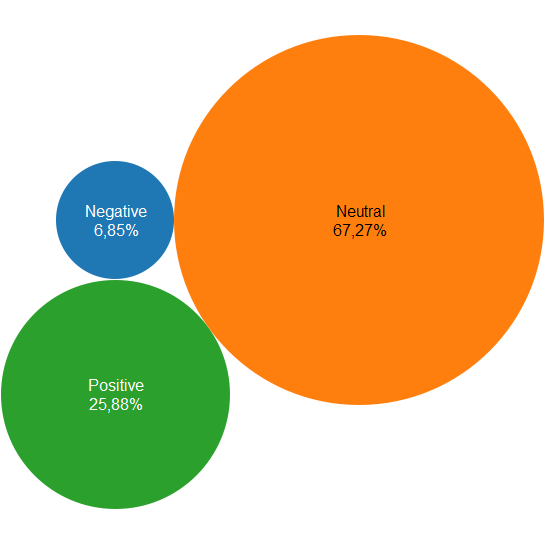}
				\caption{Twitter sentiment}
                \label{fig:twsentiment}
        \end{subfigure}
         \caption{Percentage of sentiments in Facebook and Twitter posts.}\label{fig:sentiments}
\end{figure*}

\subsection{Sentiment Analysis}

We have found positive sentiment to be more common than negative sentiment on both Facebook and Twitter. Percentages of posts with negative, positive and neutral sentiment tags are shown in Figure \ref{fig:sentiments}. Although Facebook friends are more likely to be real life acquaintances than Twitter followers, and Facebook posts are more likely to be directed to a private audience, only 22\% of Facebook posts contain a sentiment. On Twitter, tweets are public, but 32\% of tweets contain a sentiment, and 14\% are positive sentiment tweets. Negative sentiments are expressed less often on both sites, with 8\% and 6.85\% on Facebook and Twitter, respectively. In Tables \ref{table:fbsent} and \ref{table:twsent} we show some post examples with neutral, positive and negative sentiments.  

As posts on both sites are short texts, sentiments that are expressed in posts are mostly dependent on a limited number of words. For example, in Table \ref{table:twsent}, the first negative tweet has its sentiment expressed by the word `lonely'. However, the sentiment tool performs well in labeling sentiments when texts include profanity or other swear words. The following tweets are labeled  as:

\begin{itemize}
\item \textbf{Negative}: i dont care what others think any more so if you dont like me suck it up and keep your mouth shut or go f*** off. yes im in a bad mood nite

\item \textbf{Positive}: here's to me actually making some f****** money!!!!!!!!! whoo hoo!!!!!!!!!!!!!!

\item \textbf{Neutral}: F*** I, we all do.
\end{itemize}

\begin{table}[ht]
\centering
	\caption{Examples of Facebook posts with sentiments.}
	\label{table:fbsent}
	\begin{small}
  \begin{tabular}{c|l}
				
				\textbf{Positive} & I love this kid!! \\
				\cline{2-2}

				\emph{} & I like Cinebistro too! Cant wait\\
\emph{ }  & to see the Artist\\
\cline{2-2}
				\emph{} &Thanks AJ and SW crew for bringing 	\\	\emph{} & both you and Thursday back to our shores!!\\
				\hline
				\textbf{Negative} &No and no!!! \\
\cline{2-2}
				\emph{} & I dont think it worth it, it doesn't \\
\emph{}&give back as I look at it, its ugly \\
\emph{}&and... why do they call it celtic?\\	
\cline{2-2}
\emph{} & Of course it is. Their negative attacks on each \\
\emph{} & other alone are bringing some nasty skeletons\\
\emph{} & out of the closet that will hurt whichever \\
\emph{} &  Republican becomes the nominee in November.\\
\hline
\textbf{Neutral} &I never heard Shaolin monks went to Chinese \\
\emph{}&university or teach some class in school, so they\\ \emph{}&teach Chinese culture more than kungfu\\
\cline{2-2}
\emph{} &Im waiting on a special phone calll...cake up time\\
\cline{2-2}
\emph{} &They have a show in new Orleans on march 29 
\\
\end{tabular}
		\end{small} 
\end{table}

\begin{table}[ht]
\centering
	\caption{Examples of tweets with sentiments.}
	\label{table:twsent}
	\begin{small}
  \begin{tabular}{c|l}
\textbf{Positive} & watching a snuff movie, so funny, I love these things
\\
\cline{2-2}
\emph{} &hey Laura thanks for the invite.\\
\emph{} &I am eating grapes.\\
\cline{2-2}
\emph{} &Good Morning and good Ester to everyone\\
\emph{} & from Turin (Italy)!!!
\\
\hline			
\textbf{Neutral} & Signing up for twitter\\
\cline{2-2}
\emph{} &Catching up with online things...
\\
\cline{2-2}
\emph{} &Eating an apple
\\
\hline
\textbf{Negative} &so lonely =(\\
\cline{2-2}
\emph{} &lonely days.........when will these lonely \\
\emph{} & days leave me? \\
\cline{2-2}
\emph{} &is bored beyond belief
\\
\end{tabular}
		\end{small} 
\end{table}

We give the precision and recall values  for sentiment detection in Table \ref{table:sentstats}.
Overall, precision values for both online social networks are high, whereas the lowest values are obtained for positive recall for Facebook posts (54\%) and negative recall for tweets and posts (63\% and 64\%, respectively). 
These values have been obtained by asking to a group of three validators to assign a sentiment to messages, given a sample composed by 200 posts and 200 tweets. Finally, we have computed the average values of precision and recall obtained from each validator.

\begin{table}
\centering
	\caption{Precision and recall values for Twitter and Facebook sentiments. + and - signs refer to positive and negative sentiments, whereas P. and R. refer to precision and recall, respectively.}
	\label{table:sentstats}
	\begin{small}
  \begin{tabular}{c|c|c|c|c|c|c|c|c|c|c|}
				\textit{} &+P. & +R. & -P. & -R.& Neut. P. & Neut. R. \\ 
				Twitter & 88\%  & 83\%& 79\%& 63\%& 87\%& 79\%\\
				Facebook  & 81\%  & 54\%& 82\%& 64\%& 75\%& 82\%\\ 
	\end{tabular}
		\end{small} 
\end{table}

\begin{figure*}[ht]
        \centering
        \begin{subfigure}[b]{0.4\textwidth}
                \centering
                \includegraphics[scale=0.4]{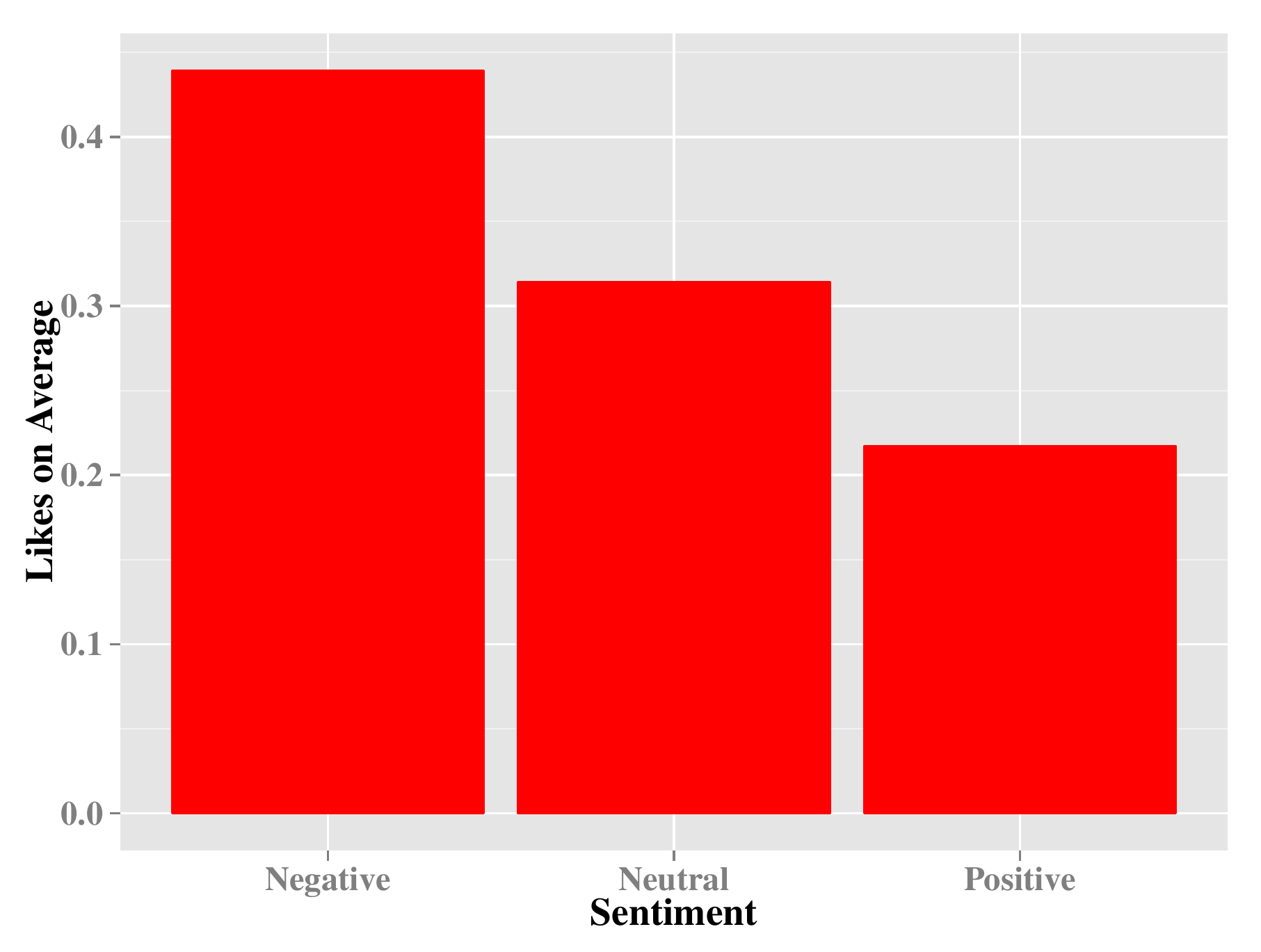}
				\caption{Likes for negative and positive comments.}
                \label{fig:retweetpos}
        \end{subfigure}%
       \begin{subfigure}[b]{0.4\textwidth}
                \centering
               \includegraphics[scale=0.4]{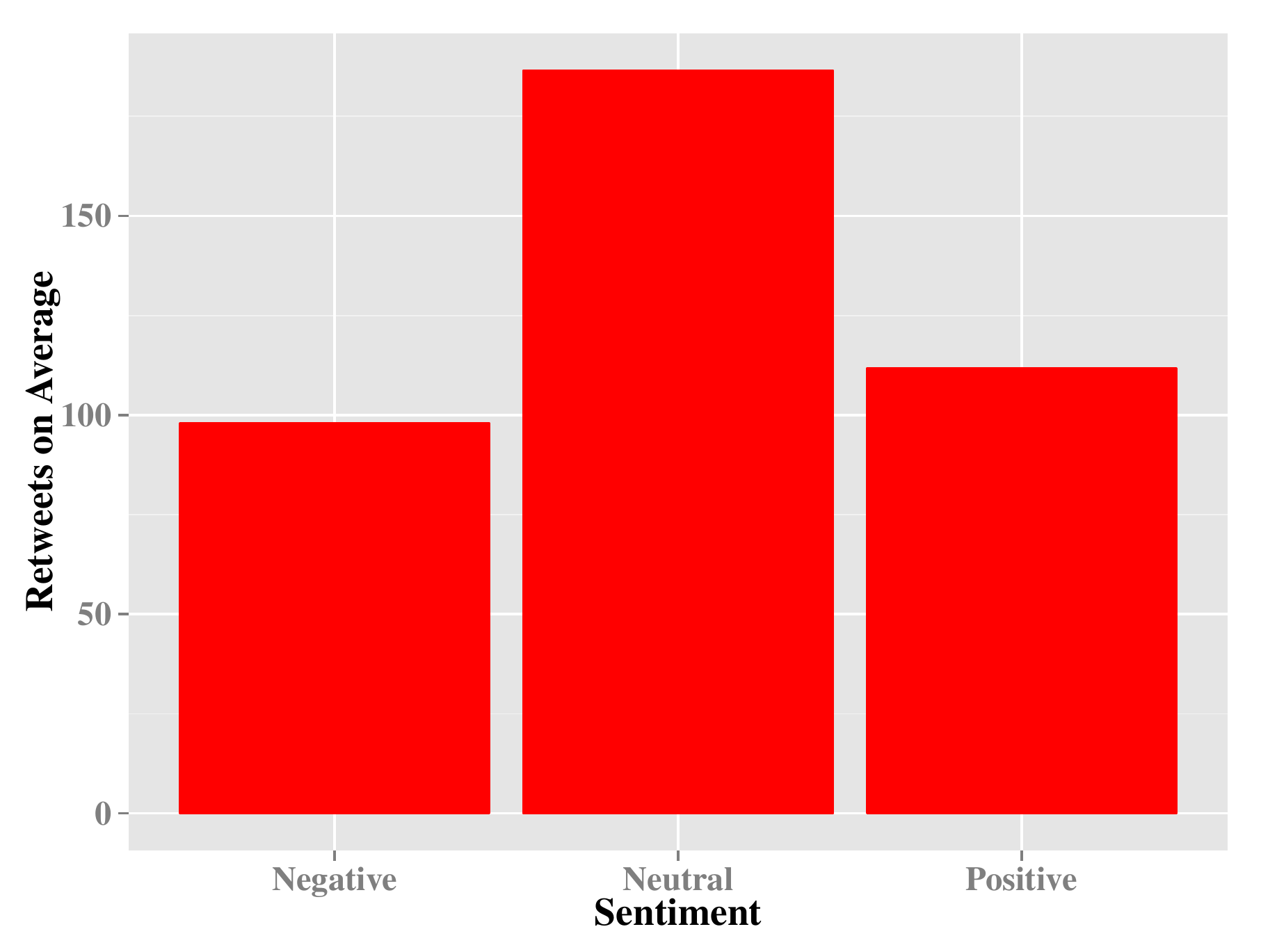}
				\caption{Retweets for negative and positive tweets.}
                \label{fig:retweetneg}
        \end{subfigure}
          \caption{The impact of sentiments on like and retweet counts. }\label{fig:percentities}
\end{figure*}

\subsection{Lexical Analysis}

In lexical analysis, we locate spelling or grammar mistakes within individual sentences of a text. To this end, we have replaced @usernames and \#hashtags on Twitter with generic words before running the lexical analysis tool, so that site specific features (e.g., hashtag usage) are stripped and the tool can parse  sentences without errors. For example, 'Ask @user about \#ny' becomes 'Ask William about New York'. With this transformation, Table \ref{table:dimstats} shows that 81\% and 69\% of Twitter and Facebook texts contain at least one lexical error.

\begin{figure}
        \includegraphics[scale=0.75]{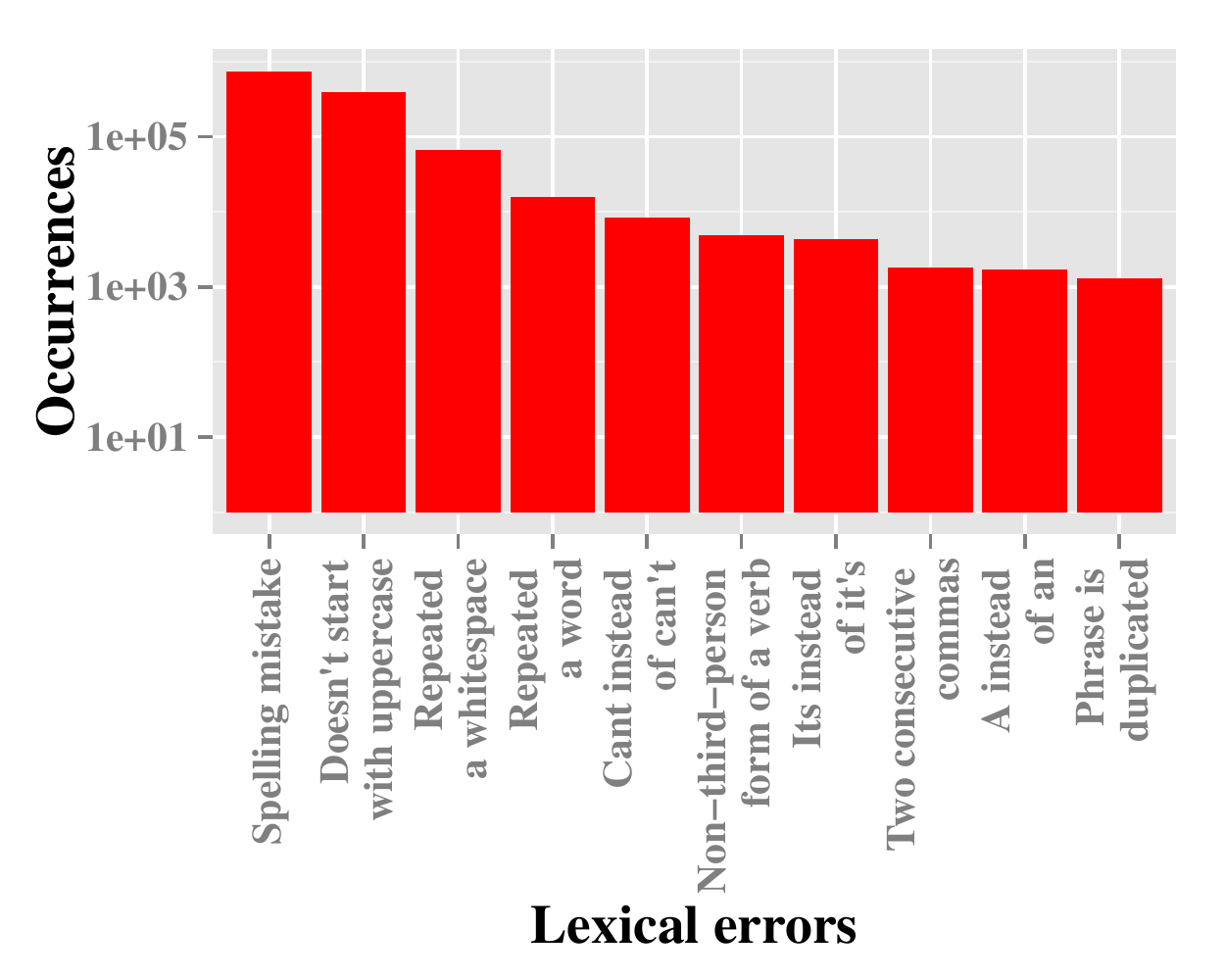}
				\caption{Most common lexical errors on Facebook.}\label{fig:fbmistake}
\end{figure}

\begin{figure}
        \includegraphics[scale=0.75]{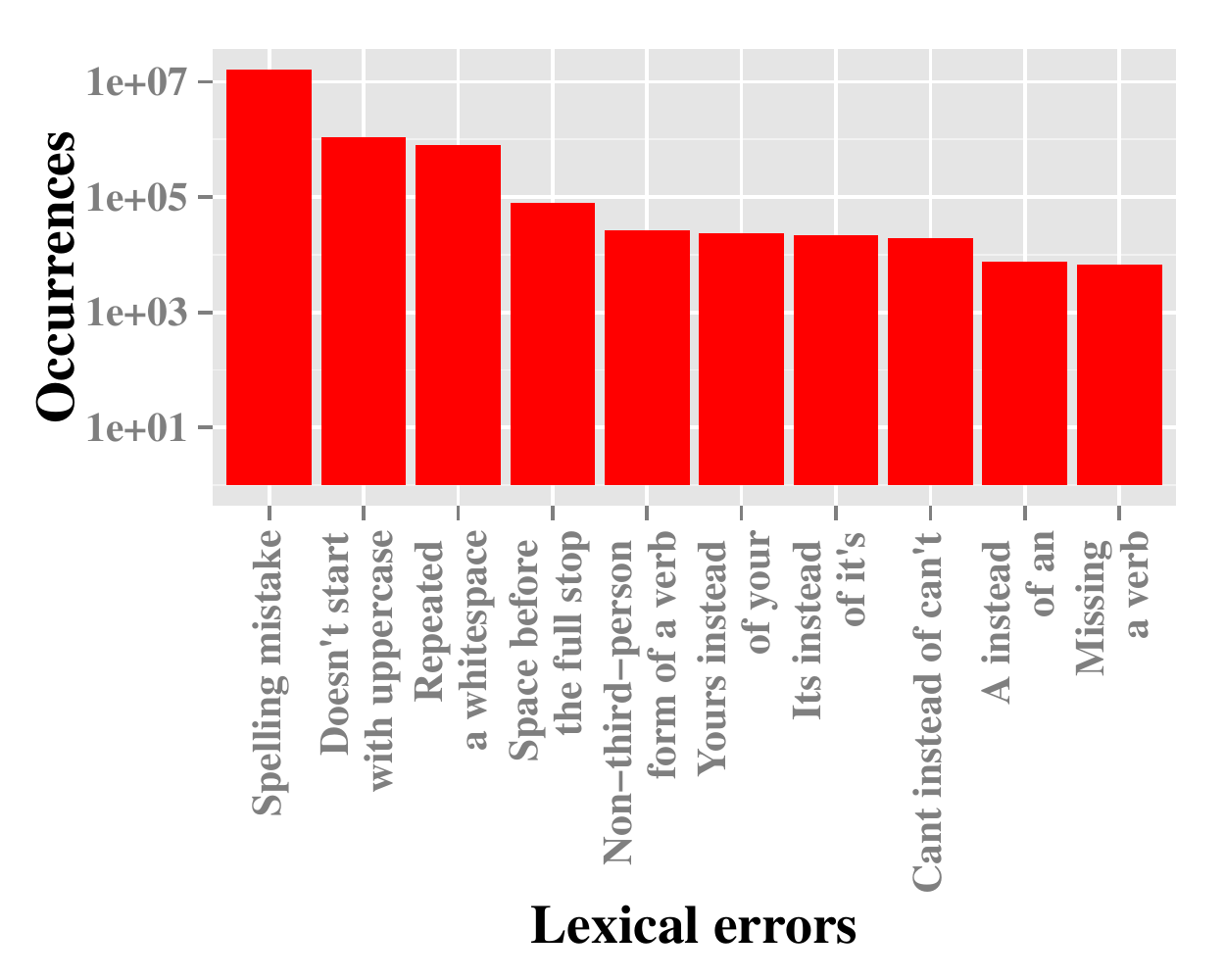}
				\caption{Most common lexical errors on Twitter.}\label{fig:twmistake}
\end{figure}

These errors are better analyzed in Figures \ref{fig:fbmistake} and \ref{fig:twmistake}.  An interesting error that appears on Twitter but not on Facebook is the absence of a proper verb in a sentence. Overall, errors on both online social networks are very similar; word spelling mistakes and absence of uppercase letters in the beginning of sentences are major errors.

\section{Dimension interplay}\label{sec:dimensioninterplay}

An interesting part of mining conversational user interactions is finding how different dimensions affect each other or retweet/like counts individually. In this section, we will look at dimension correlations and analyze how these interplays affect likes/retweets of user texts on Facebook and Twitter. In particular, we will analyze the interplay between: i) retweet/like counts and sentiment,  ii) sentiment and lexical errors, and iii) lexical errors and likes/retweets.

Figure \ref{fig:percentities} shows the impact of sentiments on retweet and like counts. In the figure, we see that negative texts receive more likes on average on Facebook, whereas on Twitter neutral tweets are retweeted more. Despite similar retweet counts, positive tweets are retweeted more often than the negative tweets.

We explain the high like count of negative Facebook posts by expressions of user compassion. Facebook posts, such as \textit{``Lexi better not die tonight. :(''} (96 likes), receive higher like counts. On Twitter neutral tweets, such as news, are retweeted by many users.

\begin{figure}
	\centering
      \includegraphics[scale=0.49]{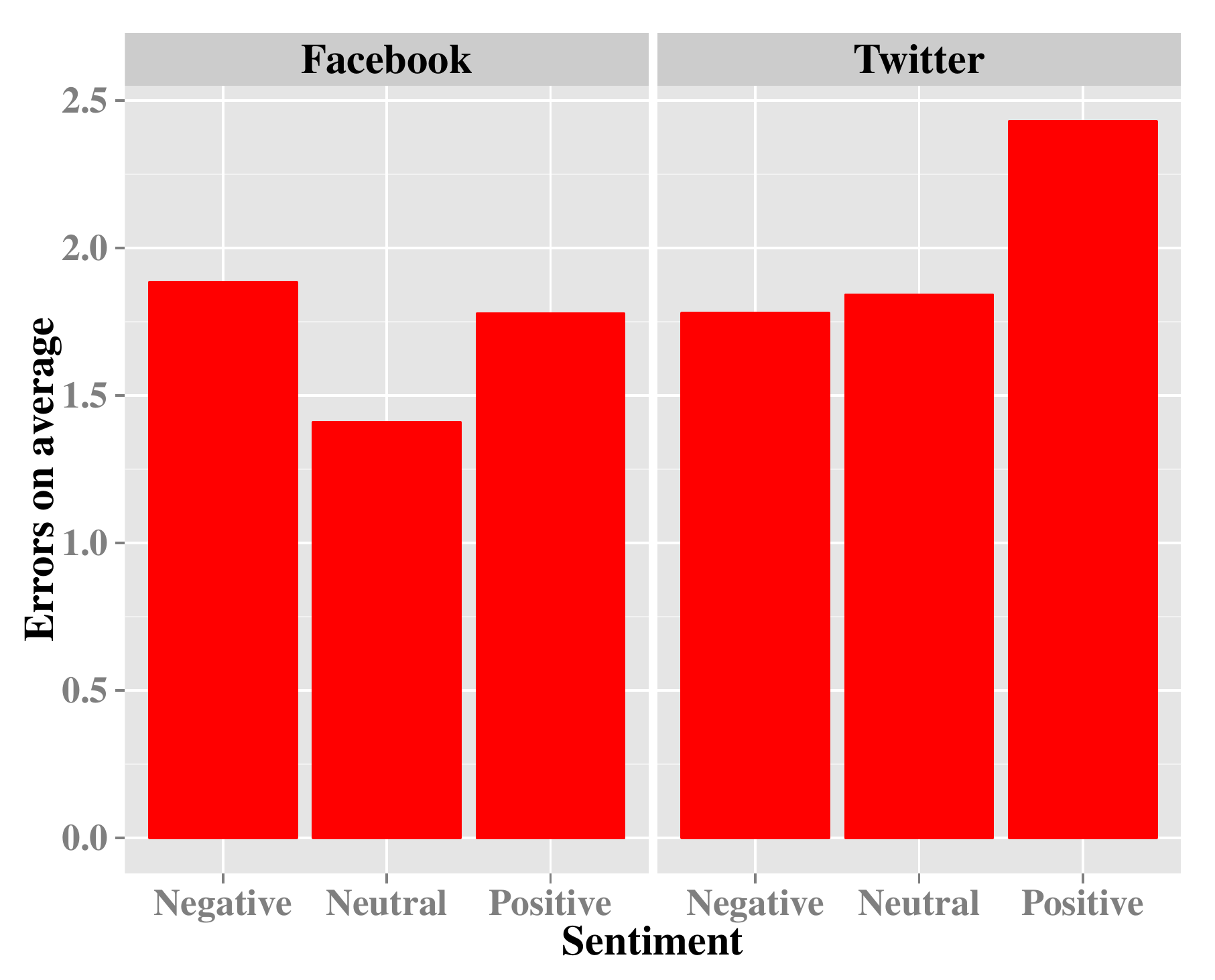}
	\caption{Average lexical errors by sentiments in Facebook and Twitter posts.}\label{fig:errsentiments}
\end{figure}

Mining both sentiments and lexical errors allows us to see how user emotions lead to lexical errors due to stress, anger or sadness.  Figure \ref{fig:errsentiments} shows average lexical errors for positive and negative sentiment texts, for both Facebook and Twitter. Positive tweets have been found to contain more than 2 errors, whereas negative Facebook comments contain 1.8 errors in average. 

In Tables \ref{table:fbsenterr} and \ref{table:twsenterr} we show some text examples with neutral, positive and negative sentiments containing errors.

\begin{table}[ht]
\centering
	\caption{Examples of Facebook posts with sentiments containing errors.}
	\label{table:fbsenterr}
	\begin{small}
  \begin{tabular}{c|l}
				
				\textbf{Positive} & strawberries would be the best.. oh my how good...\\ \emph{} & and make a chesecake from this... OH yes...\\
				\cline{2-2}
				\hline
				\textbf{Negative} &Cant believe I pay tuition money for this stuff...;-(\\
\cline{2-2}
\hline
\textbf{Neutral} &Dear NBC:  Please make your videos \\
\emph{}& viewable in other countries outside America. \\
\emph{}&Sincerly, - The rest of the world\\
\hline
\end{tabular}
		\end{small} 
\end{table}

\begin{table}[ht]
\centering
	\caption{Examples of tweets with sentiments containing errors.}
	\label{table:twsenterr}
	\begin{small}
  \begin{tabular}{c|l}
				
				\textbf{Positive} & Oh twitter! Me loves me loves! \\ \emph{} &It'll lead me to my precioussssss! \\
				\cline{2-2}
				\hline
				\textbf{Negative} &write the wrong words for the wrong thing \\
				\emph{} & and make it worse and worse. wahahha\\
\cline{2-2}
\hline
\textbf{Neutral} &looking at a little tiger cat\\
\emph{}& who says hes in my server doing maintenance \\
\hline
\end{tabular}
		\end{small} 
\end{table}

\begin{figure}
	\centering
        \includegraphics[scale=0.49]{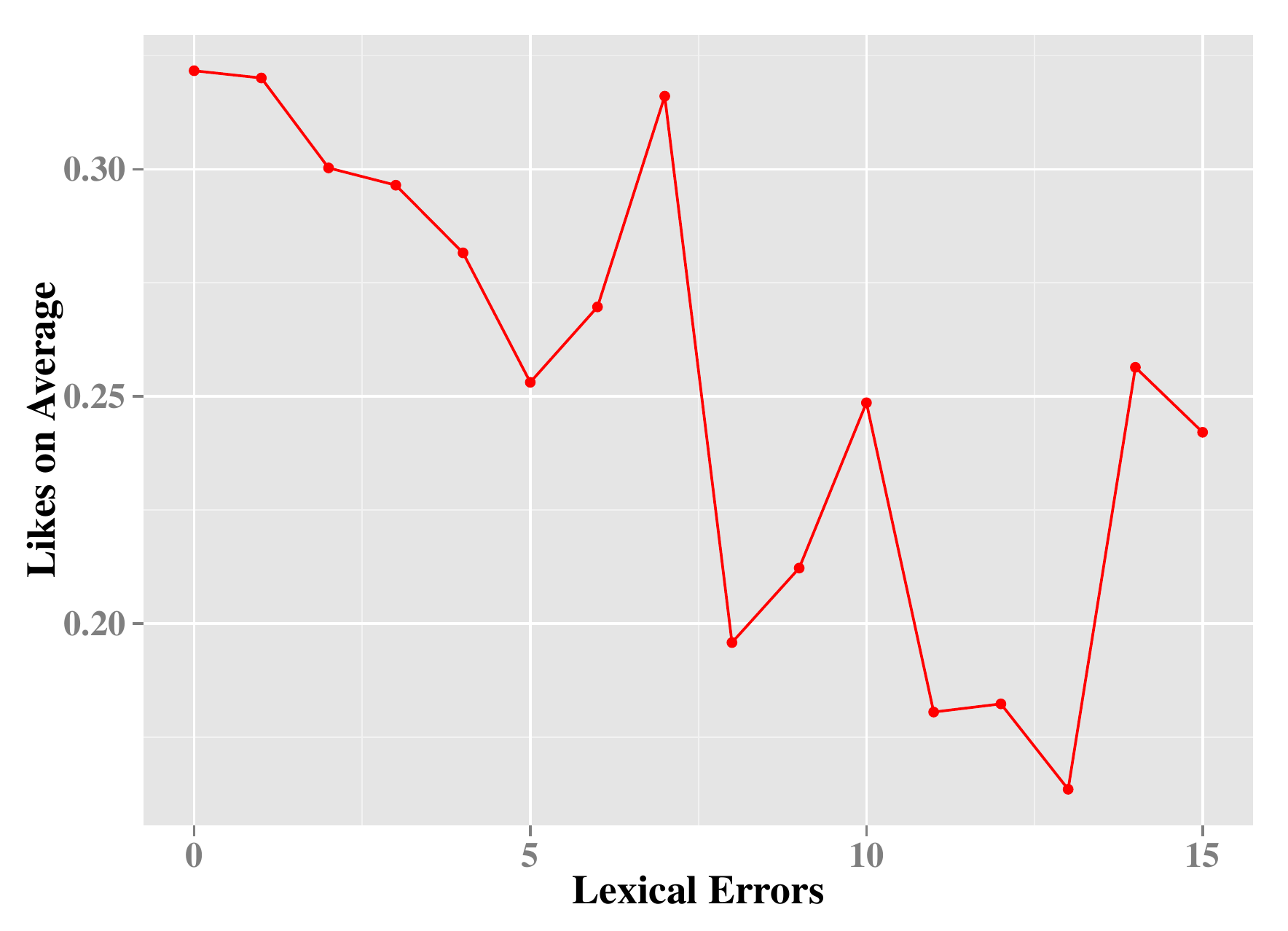}
				\caption{Average like counts by number of lexical errors.}\label{fig:likeavg}
\end{figure}

\begin{figure}
        \includegraphics[scale=0.49]{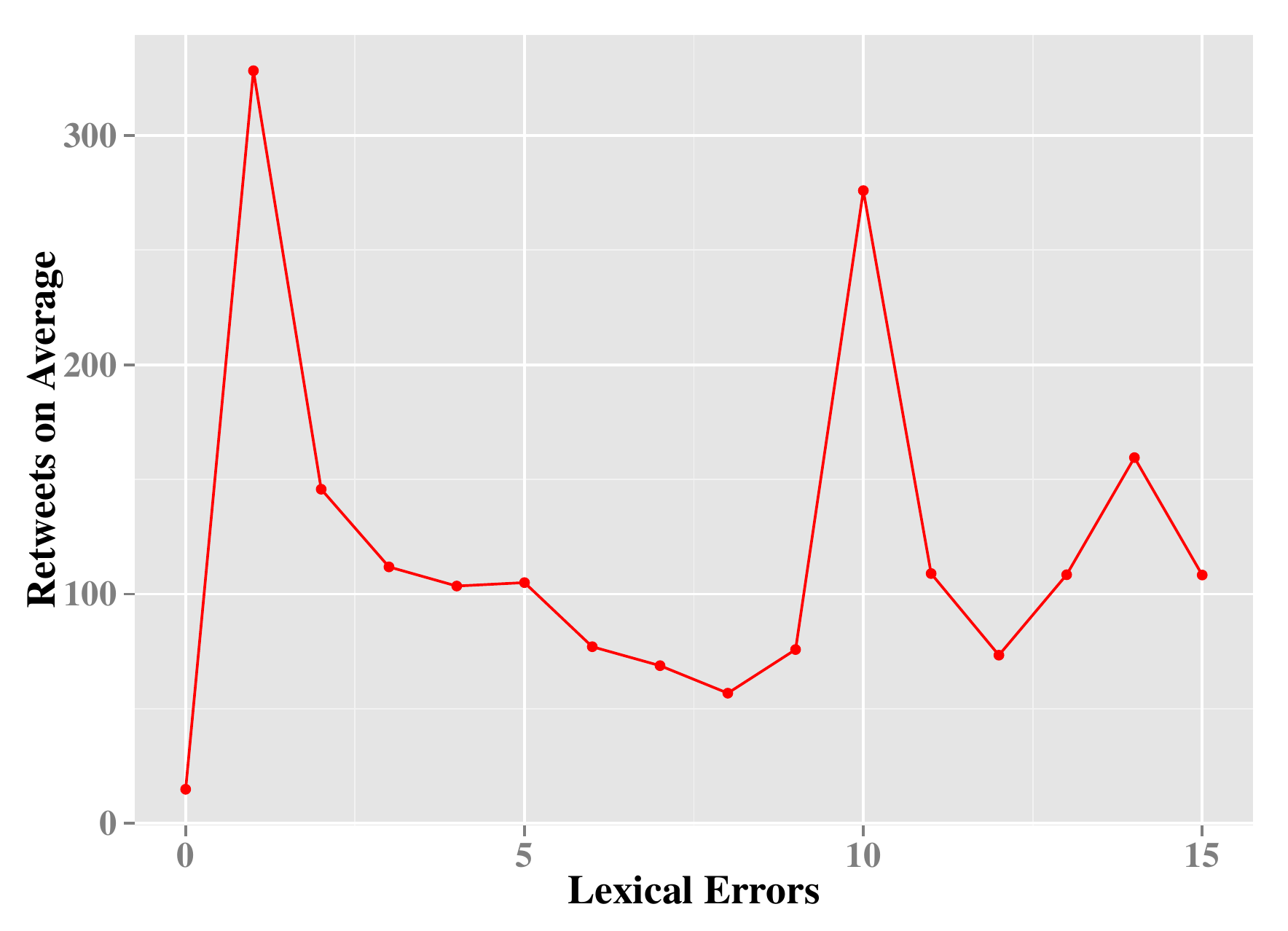}
				\caption{Average retweet counts by number of lexical errors.}\label{fig:retweetavg}
\end{figure}

Lexical errors can also affect the number of times a text is retweeted/liked by other users. In Figures \ref{fig:likeavg} and \ref{fig:retweetavg} we show the average value of likes/retweets a text receives for different values of lexical errors. Although there are as many as 15 lexical errors in both figures, number of posts with more than 5 errors are very low on both online social networks.  An example with 15 errors is the tweet \textit{omg omg omg omg omg omg omg 7 days too goo and i shall be in greece :) also my heads f***** stupid boy y does he always do this 2 me :( x}

Especially on Figure \ref{fig:likeavg} we see that an increasing number of lexical errors leads to lower like counts on Facebook. Figure \ref{fig:retweetavg} shows a similars pattern for Twitter posts, but in this case retweet counts decrease with increasing numbers of lexical errors after the first error.

\section{Influence of Geolocation on Conversations}
\label{sec:influencegeolocationconversations}

Although some research work \cite{java2007we,nagarajan2010qualitative,suh2010want} have worked on topical conversations on Twitter, a comparative analysis of locations of social network users across multiple online social networks has not been studied yet. 

\begin{figure*}[!ht]
\includegraphics[scale=0.6]{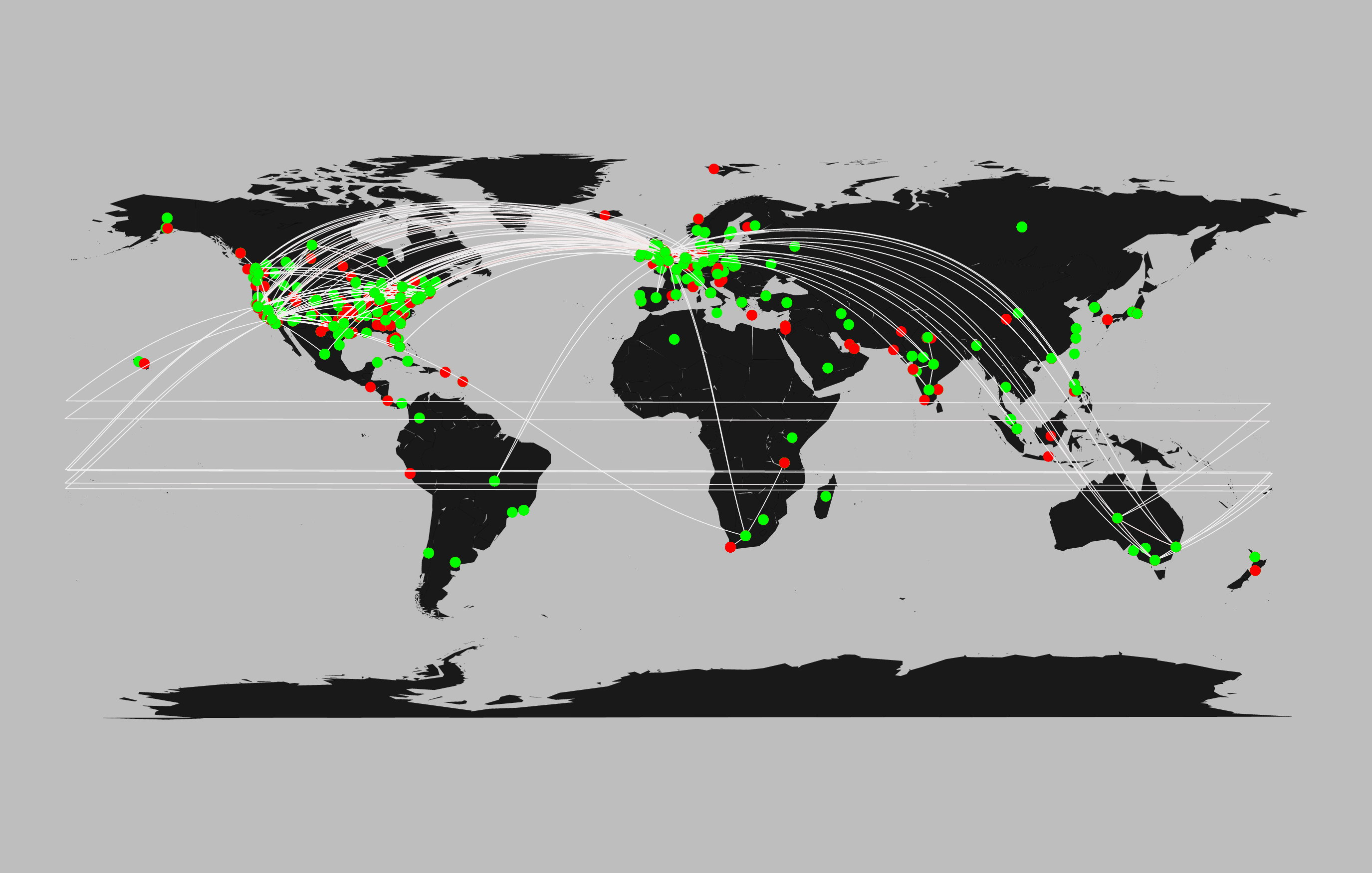}
\caption{[Color online] Locations of Twitter users. Edges show retweet behavior among Twitter users.}\label{fig:usermap}
\end{figure*}

In order to analyze the impact of geolocations, we converted textual current locations of Twitter users into geographical longitude-latitude values. From these values, Figure \ref{fig:usermap} shows the current location of Twitter users on a world map. On this map, red points and green points correspond to users who post a tweet and retweet a posted tweet, respectively. Edges between these two types of users connect them on the map. From the map, we see that a large percentage of retweeted tweets come from the east coast  of USA and north Europe. Similarly, most edges are created between these two parts of the world. The absence of China and most of the Russian territories are prominent features on the map. The high concentration of Canadian cities around the northern border of USA is also visible from the map.

\begin{figure}
        \includegraphics[scale=0.52]{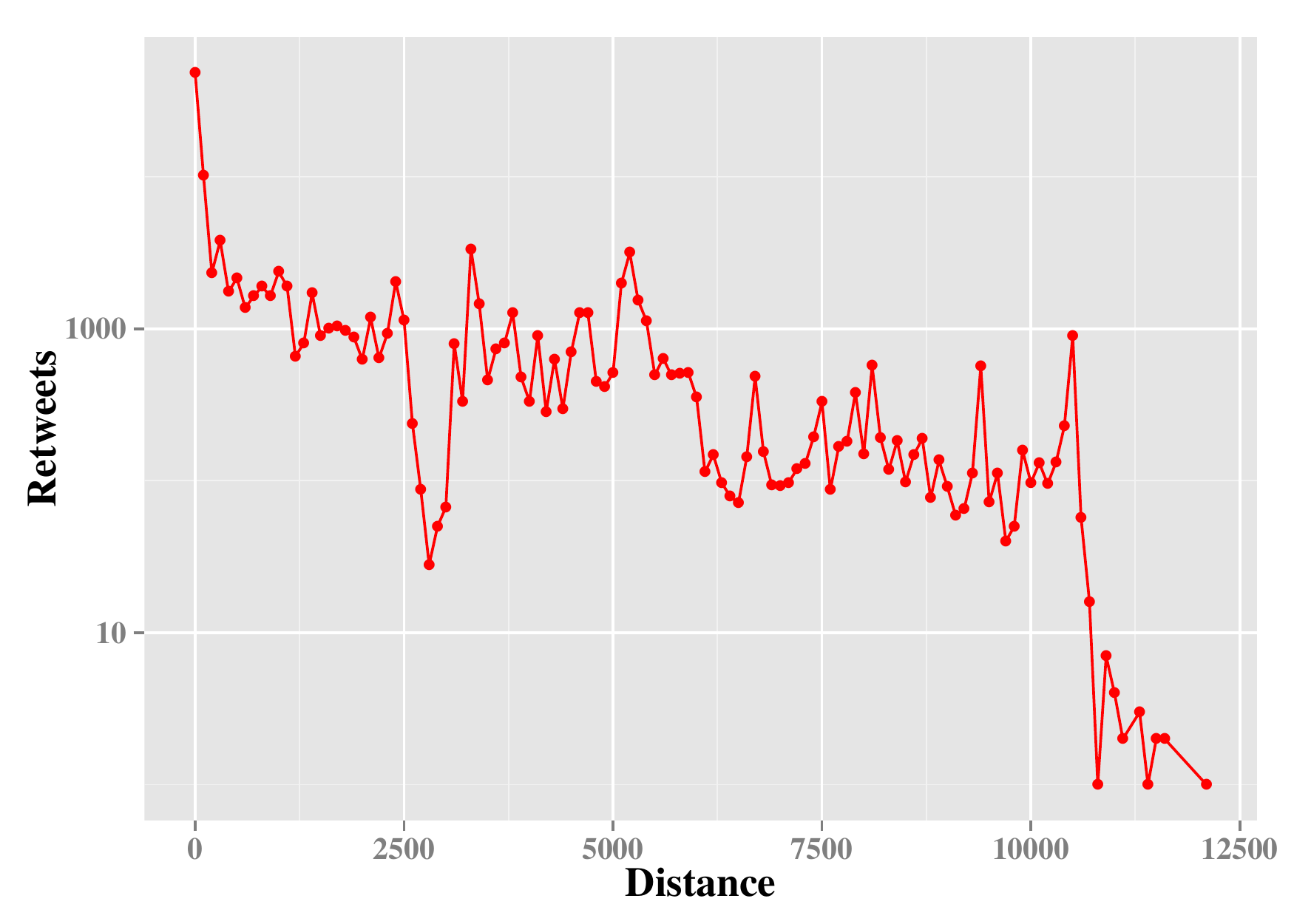}
				\caption{In miles, distances between user pairs who retweet each other's tweets. }\label{fig:distance}
\end{figure}

Another presentation of this location data allows us to measure the distance between two users in miles. A zero distance shows that a user $u$ who retweeted a tweet from user $x$ lives in the city where user $x$ currently lives. By plotting the number of such user pairs for each distinct distance value  \footnote{We have put distances into 100 mile buckets.} yields Figure \ref{fig:distance}. In the figure, we see that a big percentage of user pairs have zero distance between them. Numbers of user pairs are shown to decrease with increasing distances. An early anomaly in this trend is the low number of user pairs around 2500 miles. This distance corresponds to the separation between USA and Europe.

Unlike Twitter, Facebook privacy settings are very strict, and users are less willing to share their location information. Due to this shortcoming, we could not repeat the distance experiment on the Facebook dataset. 
For a similar distance notion on Facebook dataset, we have used the locale \footnote{Locale is the user chosen interface language for Facebook.com, such as EN\_US (English in USA), EN\_GB (English in Great Britain)} field to analyze user proximity.  We explain this choice with the empirical observation that users from a country mainly use the official language of the country in the  Facebook interface. The main exception to this observation is that English is also widely used by other nationalities.

\begin{figure}
        \includegraphics[scale=0.4]{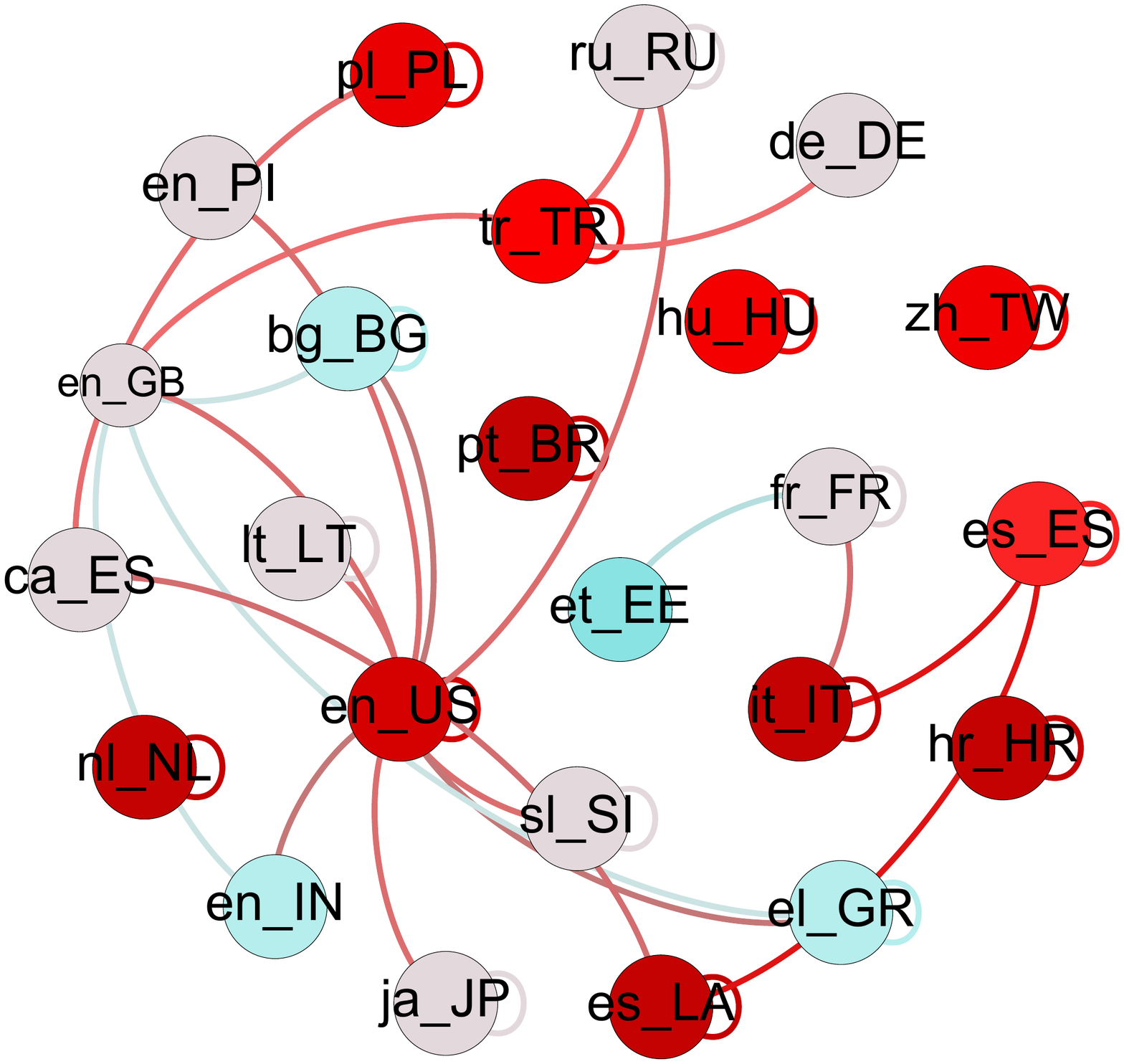}
				\caption{Like interactions among  Facebook users of different locales. Self loops show that most of users from a locale likes comments from users of the same locale. }\label{fig:locale}
\end{figure}

Figure \ref{fig:locale} shows locale pairs for Facebook users. In the figure, a locale value $l$ is connected by an edge to another locale $m$ if users from $l$ constitute 30\% or more of users who have liked Facebook comments of users from the locale $m$. Although EN\_US is connected to a big portion of other locales, most locales have self loops, or they are connected to a small number of other locales. Another relevant feature of the figure is the community of Latin languages connected together, such as it\_IT, es\_ES, fr\_FR, etc.; users of Latin languages interact more often.

In the previous sections, we have shown how different dimensions of user generated texts affect how many times a particular tweet/comment will be liked or commented. Regardless of these numbers,  our geolocation experiments show that  users who like/comment a text are more likely to live closer to the owner of the text.

\section{Conclusion}

In this paper, we have analyzed conversational user interactions on two popular online social networks, Facebook and Twitter. We have found common user behavior in interacting with posts of similar sentiments (i.e., positive or negative). Furthermore, texts on both web sites have been found to exhibit similar lexical errors, but these errors result in differing behaviors in user interactions.

Our conversational analysis has been complemented with a location analysis of users on both online social networks. This  approach has shown that geolocations of users can greatly affect which social network posts will be liked and retweeted and will have better visibility.

From an information propagation point of view, our results can help in choosing seed nodes to disseminate news or advertisements effectively. Furthermore, propagation of any data can greatly benefit from an increased location awareness, because users tend to interact with others who are from the same locations.

\bibliographystyle{plain}	
\bibliography{ref}		

\end{document}